# Chipscale plasmonic modulators and switches based on metal-insulator-metal waveguides with Ge$_2$Sb$_2$Te$_5$

Zhaojian Zhang [1], Junbo Yang [2, *], Wei Bai [3], Yunxin Han [2], Xin He [2], Jingjing Zhang [1], Jie Huang [1], Dingbo Chen [1], Siyu Xu [1] and Wanlin Xie [1]

[1] College of Liberal Arts and Sciences, National University of Defense Technology, Changsha 410073, China

[2] Center of Material Science, National University of Defense Technology, Changsha 410073, China

[3] Institute of Optics and Electronics Chinese Academy of Sciences, Chengdu 610209, China

*Correspondence: yangjunbo@nudt.edu.cn

**Abstract:** We introduce phase-change material Ge$_2$Sb$_2$Te$_5$ (GST) into metal-insulator-metal (MIM) waveguide systems to realize chipscale plasmonic modulators and switches in the telecommunication band. Benefitting from the high contrast of optical properties between amorphous and crystalline GST, the three proposed structures can act as reconfigurable and non-volatile modulators and switches with excellent modulation depth 14 dB and fast response time in nanosecond, meanwhile possessing small footprints, simple frameworks and easy fabrication. This work provides new solutions to design active devices in MIM waveguide systems, and can find potential applications in more compact all-optical circuits for information processing and storage.

**1. Introduction**

Facing the increasing demand for information quantity and transmission rate, today's electronic integrated devices will be unable to afford roles in future chipscale systems due to the fundamental limits [1]. As a promising solution, photonic integrated devices can break the blocks by using light as the carrier of information. To construct basic photonic circuits, dielectric waveguides (such as silicon waveguide) have been widely studied for decades [2,3]. However, the scale of a dielectric waveguide cannot go smaller than the working wavelength due to the diffraction limit, which restricts the integration and compactness of photonic integrated chips [1]. Recently, surface plasmon polariton (SPPs), which are the surface electromagnetic waves supported at the interface between metal and dielectric, have attracted more attention due to the capacity to break the diffraction limit [1]. SPPs also introduce plasmonic waveguides, which can deliver light in sub-wavelength scale [4,5]. Especially, plasmonic metal-insulator-metal (MIM) waveguides possess good confinement of light, low bending loss, easy fabrication and acceptable propagation length, therefore can be potential platforms for future more compact all-optical integrated circuits [6-9].

As the promising next-generation on-chip photonic system, MIM system has supported various chipscale passive devices such as filters [10,11], demultiplexers [12,13], sensors [14,15] and spacers [16,17]. However, active devices are also required in such system. Therefore, optical materials with tunable properties have been utilized to realize dynamically tunable functionalities. For example, nonlinear materials have been used to realize all-optical tunable filters, logic gates and diodes [18-20]. Lasers, switches and slow light enhancement can be achieved by integrating gain materials [21-23]. Two-dimension (2D) materials have contributed to the active manipulation of MIM waveguides as well [24-26]. Some novel optical concepts, such as epsilon-near-zero (ENZ) effects and non-PT-symmetry, have also been made possible in MIM systems assisted by optical materials [27,28]. Obviously, optical materials will play an important role in active systems, and it is essential to explore more outcomes from the combination between MIM waveguides and novel optical materials. Recently, phase-change materials (PCMs), such as $Ge_2Sb_2Te_5$ (GST), have drawn more attention. GST possesses tremendously different optical properties from amorphous to crystalline states. Such phase transition can be triggered by thermal, electrical or optical schemes, has an ultrafast switching time in nanosecond (ns) or even sub-ns, and is usually reversible [29]. Up to now, GST has been regarded as a strong candidate for realizing reconfigurable and non-volatile all-optical devices for data processing and storage [30], and many applications have been proposed such as on-chip memory elements [31-33], all-optical switching [34-36] and reconfigurable metasurfaces [37-39].

Here, we introduce GST into MIM systems to realize chipscale plasmonic modulators and switches in shortwave infrared (SWIR) telecommunication regime. Combining metal-silicon-metal (MSM) waveguides and metal-GST-metal (MGM) resonators, we numerically propose three typical MIM structures, including end-coupled rectangular resonator, side-coupled stub resonator and two mutually coupled resonators, to investigate their performances on transmission modulations. Due to the phase transition of GST, the effective refractive index inside MGM resonators can be actively tuned with high contrast, consequently leading to the significant shifts of resonant wavelengths. Benefitting from this, the three proposed structures can act as reconfigurable and non-volatile modulators and switches with excellent modulation depth and fast response time, meanwhile possessing small footprints, simple frameworks and easy fabrication. This work provides new solutions to design active devices based on MIM systems, and can find potential applications in future on-chip all-optical circuits.

## 2. Structures, materials and methods

The first proposed structure is depicted in Fig. 1, consisting of input and output MSM waveguides end-coupled with an MGM rectangular resonator. The geometric parameters are as follows: the width of waveguide $w$= 50 nm, the gap $g$= 20 nm, and the length of resonator $L$= 120 nm.

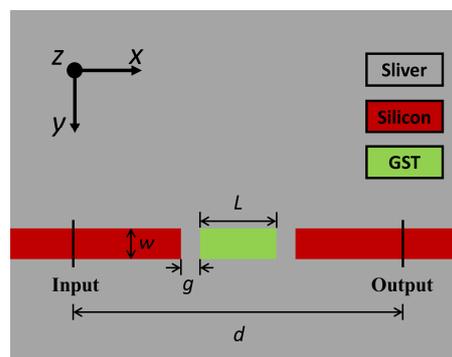

Fig. 1 The schematic of the first structure.

In Fig. 1, the grey area is silver (Ag), the dielectric constant of which is defined by the Drude model [10]:

$$\varepsilon_m = \varepsilon_\infty - \frac{\omega_p^2}{\omega(\omega+i\gamma)} \qquad (1)$$

Where $\varepsilon_m$ is the dielectric constant of Ag, $\varepsilon_\infty$ is the dielectric constant at infinite frequency, $\omega_p$ is the bulk plasma frequency, $\gamma$ is the electron oscillation damping frequency, and $\omega$ is the angular frequencies of incident waves. The parameters of Ag are as follows: $\varepsilon_\infty = 3.7$, $\omega_p = 1.38\times10^{16}$ Hz, and $\gamma = 2.73\times10^{13}$ Hz. The waveguides are filled with silicon (Si) due to the low dispersion and loss in SWIR which is shown in red in Fig. 1, and the corresponding optical constant ($n+ik$) are presented in Fig. 2 [40]. GST possesses dramatically different optical properties between the amorphous (aGST) and crystalline state (cGST), the corresponding optical constants are also shown in Fig. 2, respectively [41]. We can see that there is a high contrast in refractive index between the two states, and cGST possesses much more intrinsic loss than aGST in SWIR. The effective dielectric constants of GST with different crystallinities are predicted by the Lorentz-Lorenz relation [42]:

$$\frac{\varepsilon_{GST}(\lambda,C)-1}{\varepsilon_{GST}(\lambda,C)-1} = C\times\frac{\varepsilon_{cGST}(\lambda)-1}{\varepsilon_{cGST}(\lambda)+2} + (1-C)\times\frac{\varepsilon_{aGST}(\lambda)-1}{\varepsilon_{aGST}(\lambda)+2} \qquad (2)$$

Here, $\varepsilon_{aGST}$ and $\varepsilon_{cGST}$ are dielectric constants of aGST and cGST, respectively. $C$ is the crystallization fraction of GST from 0 to 1.

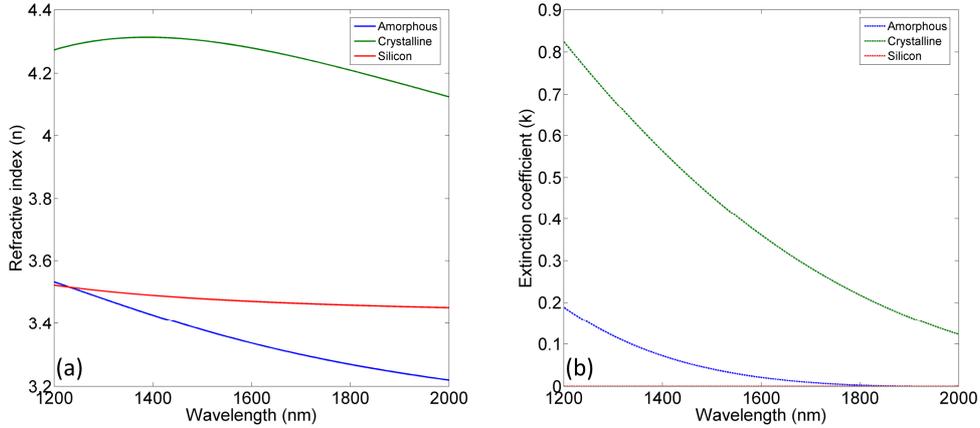

Fig. 2 (a) The refractive index of amorphous GST, crystalline GST and silicon. (b) The extinction coefficient of amorphous GST, crystalline GST and silicon.

In the MIM waveguide, only fundamental transverse-magnetic (TM) mode exists. The dispersion of fundamental mode is described as follows [6]:

$$\frac{\varepsilon_i\sqrt{\beta^2-\varepsilon_m k_0^2}}{\varepsilon_m\sqrt{\beta^2-\varepsilon_i k_0^2}} + \tanh(\frac{w\sqrt{\beta^2-\varepsilon_i k_0^2}}{2}) = 0 \qquad (3)$$

Where $\varepsilon_i$ is the dielectric constant of the insulator (Si or GST), $k_0 = 2\pi/\lambda_0$ represents the wavevector in free space, and $\beta$ is the wavevector of SPPs inside the MIM waveguide. The effective refractive index is defined as $n_{eff} = \beta/k_0$. The calculated $n_{eff}$ of fundamental TM mode in the MSM waveguide as well as the MGM waveguide in different crystallization conditions are shown in Fig. 3(a) and (b). The profile of fundamental TM mode in the MSM waveguide is presented in Fig. 3(c), and the mode in the MGM waveguide has the similar profile. 2D Finite-Difference Time-Domain (FDTD) method is applied to do the simulation, the mesh size is set as 3 nm to ensure the accuracy, and the boundary condition is set as perfectly matched layers (PMLs) to maintain convergence. Two power monitors are placed at input and output respectively, and the distance between the two monitors is $d$= 500 nm. The transmission is calculated following $T= P_{out}/P_{in}$.

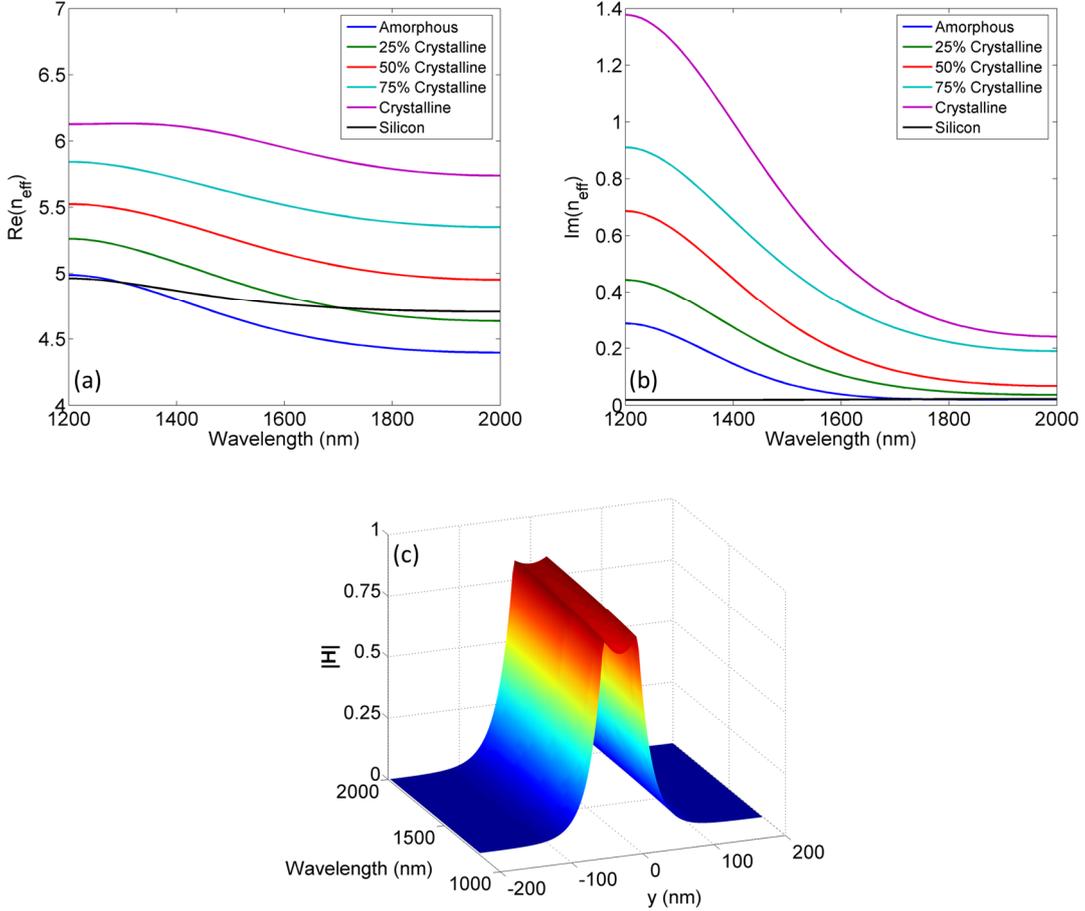

Fig. 3 (a) The real part of effective refractive index. (b) The imaginary part of effective refractive index. (c) The profile of fundamental TM mode in the MSM waveguide.

## 3. Results and discussions

The transmission spectrum of the structure in Fig. 1 corresponding to different phase states of GST are shown in Fig. 4(a). In MIM system, such structure is a typical band-pass filter. The rectangular resonator acts as a Fabry–Perot (F-P) resonator, only resonant wavelength can be coupled into this resonator from the input waveguide, and released to the output waveguide. The resonant condition is as follows [43]:

$$(2m - \frac{\varphi}{\pi})\lambda = 4\operatorname{Re}(n_{eff})L, \ m = 1, 2... \quad (4)$$

Here, $m$ is the mode number, which is an integer. $\varphi$ is the phase shift induced by the reflection at the insulator-metal interface. As shown in Fig. 4(a) in blue, the passed resonant wavelength is 1527 nm when GST is in amorphous state, the transmission of which is 44%. The corresponding magnitude of magnetic field |$H$| distribution is shown in Fig. 4(b), from which we can see that there exists a F-P mode with $m$= 1 in the rectangular resonator, and SPPs can pass through such structure. However, When GST comes to crystalline state, the transmission peak will have a red shift due to the increase of $\text{Re}(n_{eff})$ as shown in Fig. 4(a). Meanwhile, the amplitude of the peak will decline caused by the rise of material loss. Benefitting from this, the transmission at 1527 nm can be actively tuned from 44% to 1.8%, the corresponding |*H*| distribution is shown in Fig. 4(c), showing that transmitted SPPs is suppressed. To evaluate the performance of transmission modulation, the modulation depth (MD) is defined as follows:

$$\text{MD} = \frac{T_{max} - T_{min}}{T_{max}} \times 100\% \qquad (5)$$

Where $T_{max}$ and $T_{min}$ are maximum and minimum transmission values, respectively. For convenience, we also provide modulation results in log scale:

$$\begin{aligned} \text{IL (Insertion loss)} &= -10\lg(T) \\ \text{MD}_{dB} &= -10\lg(T_{min}/T_{max}) \end{aligned} \qquad (6)$$

The corresponding modulation results are shown in Table. 1. From the aGST to cGST, the MD can reach 96% (13.8 dB). Such modulation is much better than previous similar works based on MIM structures [44-48], meanwhile possesses comparable response time.

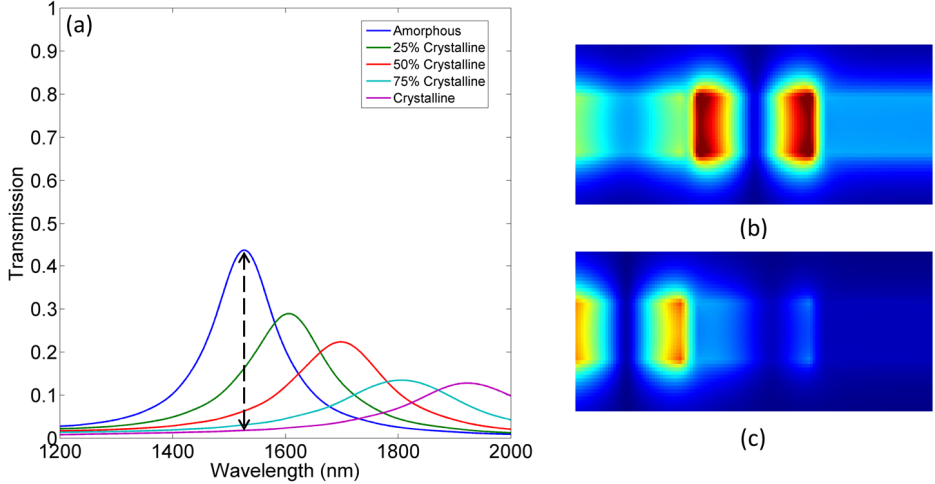

Fig. 4 (a) The transmission spectrum corresponding to different phase states of GST. (b) The |*H*| distribution at 1527 nm in GST amorphous state. (c) The |*H*| distribution at 1527 nm in GST crystalline state.

Table 1. The transmission as well as insertion loss in different GST states and the modulation depth.

| GST state | a | 25% c | 50% c | 75% c | c | MD |
|---|---|---|---|---|---|---|
| T | 44% | 16% | 6.3% | 3.1% | 1.8% | 96% |
| IL | 3.6 dB | 8.0 dB | 12.0 dB | 15.1 dB | 17.4 dB | 13.8 dB |

Next, we introduce the second structure, a main MSM waveguide side-coupled with an MGM stub resonator, as shown in Fig. 5. The geometric parameters are as follows: *w*= 50 nm and *L*= 200 nm. The other parameters are unchanged.

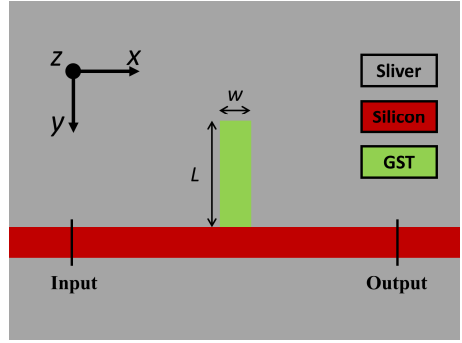

Fig. 5 The schematic of the second structure.

Such structure is a typical band-stop filter in MIM system. The stub resonator can be seen as a semi F-P resonator, can trap the resonant wavelength and prevent it from passing, the resonant condition is as follows [49]:

$$(2n+1-\frac{\varphi}{\pi})\lambda = 4\operatorname{Re}(n_{eff})L, \ n=0,1... \quad (7)$$

Where *n* is the mode number. The transmission spectrum corresponding to different phase states of GST are shown in Fig. 6(a). Due to the high contrast between the refractive index of aGST and cGST, a giant red shift of transmission dip happens. Consequently, the transmission at 1513 nm can be modulated from 2.1% (16.8 dB) to 57% (2.4 dB), leading to an excellent MD 96% (14.4 dB), the details are given in table. 2. The |***H***| distribution at 1513 nm corresponding to aGST and cGST are depicted in Fig. 6 (b) and (c), respectively, indicating *n*= 1.

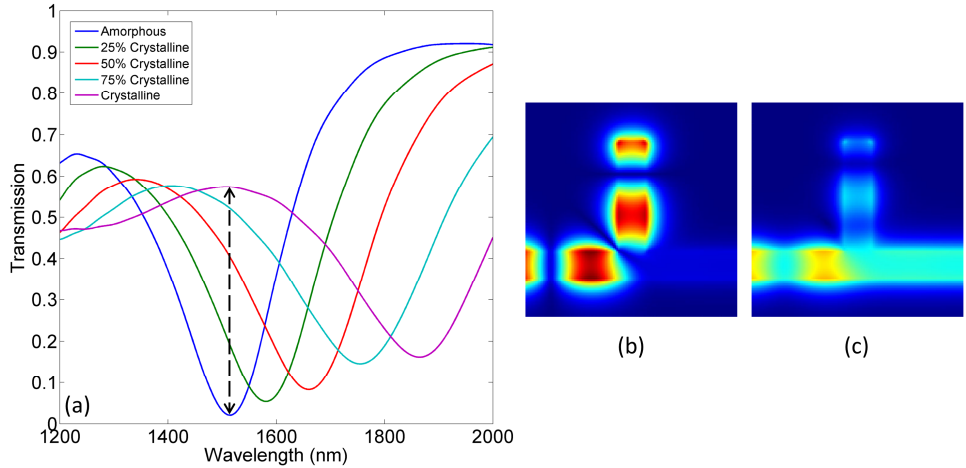

Fig. 6 (a) The transmission spectrum corresponding to different phase states of GST. (b) The |***H***| distribution at 1513 nm in GST amorphous state. (c) The |***H***| distribution at 1513 nm in GST crystalline state.

Table 2. The transmission as well as insertion loss in different GST states and the modulation depth.

| GST state | a | 25% c | 50% c | 75% c | c | MD |
|---|---|---|---|---|---|---|
| T | 2.1% | 19% | 41% | 52% | 57% | 96% |
| IL | 16.8 dB | 7.2 dB | 3.9 dB | 2.8 dB | 2.4 dB | 14.4 dB |

The last structure is shown in Fig. 7, including a main MSM waveguide side-coupled with a stub resonator and a rectangular resonator. The geometric parameters are as follows: *h*= 45 nm, *g*= 15 nm, *L*= 110 nm and *w*= 50 nm. The other parameters are unchanged.

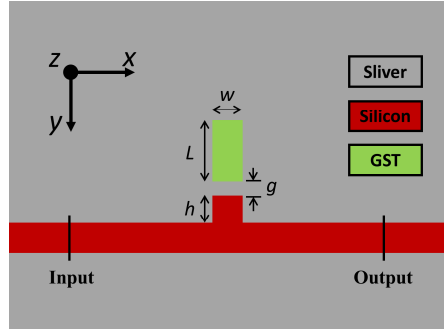

Fig. 7 The schematic of the third structure.

Such structure can produce plasmon-induced transparency (PIT) effect, which is a special case of Fano resonance. When this is only stub resonator, a transmission dip exists as shown in Fig. 8(a) in blue. However, when the rectangular resonator is coupled upside the stub resonator, a transmission peak will appear at the position of the original dip as given in Fig. 8(a) in green, which is called PIT. Such effect attributes to the near-field interaction between the mode in the stub (radiative mode) and the mode in the rectangle (subradiant mode). When the two modes have close resonant wavelengths, constructive interference will happen between the two modes, consequently leading to that transparent window [50].

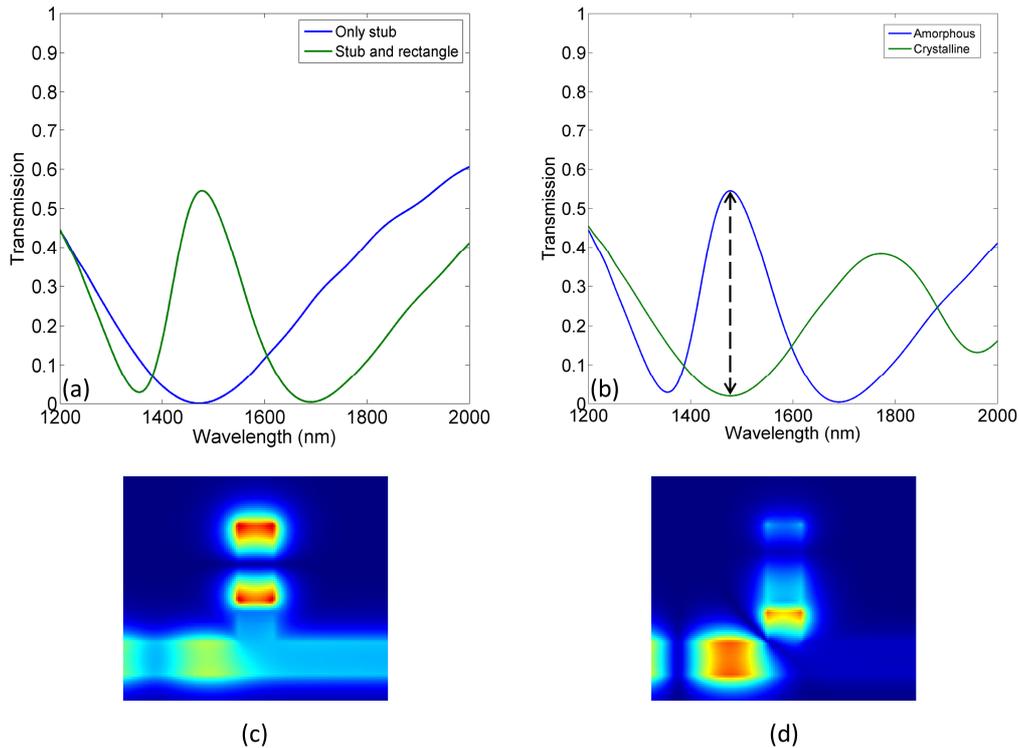

Fig. 8 (a) The transmission spectrum corresponding to only stub as well as stub plus rectangle. (b) The transmission spectrum corresponding to aGST and cGST. (c) The |$H$| distribution at 1479 nm in aGST. (d) The |$H$| distribution at 1479 nm in cGST.

Here, we utilize GST to tune the resonant wavelengths of the subradiant mode, and further control the interference to realize switching. When GST is amorphous, the resonant wavelengths of radiative and subradiant modes are close to each other, therefore leading to an induced transparent peak at 1479 nm as shown in Fig. 8 (b) in blue. However, when it comes to cGST, the resonant wavelength of subradiant mode is shifted away, leading to no interference i.e., a transmission dip at 1479 nm as shown in Fig. 8(b) in green. The corresponding |$H$|

distributions are given in Fig. 8(c) and (d), respectively. The transmission at 1479 nm can be switched from 55% (2.6 dB) to 2% (17.0 dB), the MD can be up to 96% (14.4 dB), which is promising for all-optical switching.

## 4. Conclusion

In summary, we investigate three structures to show that, GST can be efficiently utilized to construct the high-performance modulator and switch in the plasmonic MIM system, and it is the first time to introduce GST into MIM system so far as we know. Besides, this work focuses on telecommunication band, aiming to provide more solutions for on-chip photonic systems for information processing. Furthermore, the small footprint and easy fabrication of proposed structures can have advantages in applications on future more compact all-optical circuits.

**Acknowledgments**

This work is supported by the National Natural Science Foundation of China (61671455, 61805278), the Foundation of NUDT (ZK17-03-01), the Program for New Century Excellent Talents in University (NCET-12-0142), and the China Postdoctoral Science Foundation (2018M633704).

**Reference**

[1] Gramotnev, D. K., & Bozhevolnyi, S. I. (2010). Plasmonics beyond the diffraction limit. Nature photonics, 4(2), 83.

[2] Marcatili, E. A. (1969). Dielectric rectangular waveguide and directional coupler for integrated optics. Bell System Technical Journal, 48(7), 2071-2102.

[3] Deslandes, D., & Wu, K. (2001). Integrated microstrip and rectangular waveguide in planar form. IEEE Microwave and Wireless Components Letters, 11(2), 68-70.

[4] Bozhevolnyi, S. I., Volkov, V. S., Devaux, E., Laluet, J. Y., & Ebbesen, T. W. (2006). Channel plasmon subwavelength waveguide components including interferometers and ring resonators. Nature, 440(7083), 508.

[5] Fang, Y., & Sun, M. (2015). Nanoplasmonic waveguides: towards applications in integrated nanophotonic circuits. Light: Science & Applications, 4(6), e294.

[6] Dionne, J. A., Sweatlock, L. A., Atwater, H. A., & Polman, A. (2006). Plasmon slot waveguides: Towards chip-scale propagation with subwavelength-scale localization. Physical Review B, 73(3), 035407.

[7] Neutens, P., Van Dorpe, P., De Vlaminck, I., Lagae, L., & Borghs, G. (2009). Electrical detection of confined gap plasmons in metal–insulator–metal waveguides. Nature Photonics, 3(5), 283.

[8] Lu, H., Wang, G., & Liu, X. (2013). Manipulation of light in MIM plasmonic waveguide systems. Chinese Science Bulletin, 58(30), 3607-3616.

[9] Kriesch, A., Burgos, S. P., Ploss, D., Pfeifer, H., Atwater, H. A., & Peschel, U. (2013). Functional plasmonic nanocircuits with low insertion and propagation losses. Nano letters, 13(9), 4539-4545.

[10] Lu, H., Liu, X., Mao, D., Wang, L., & Gong, Y. (2010). Tunable band-pass plasmonic waveguide filters with nanodisk resonators. Optics Express, 18(17), 17922-17927.

[11] Zhan, G., Liang, R., Liang, H., Luo, J., & Zhao, R. (2014). Asymmetric band-pass plasmonic nanodisk filter with mode inhibition and spectrally splitting capabilities. Optics express, 22(8), 9912-9919.

[12] Wang, G., Lu, H., Liu, X., Mao, D., & Duan, L. (2011). Tunable multi-channel wavelength demultiplexer based on MIM plasmonic nanodisk resonators at telecommunication regime. Optics Express, 19(4), 3513-3518.

[13] Nakayama, K., Tonooka, Y., Ota, M., Ishii, Y., & Fukuda, M. (2018). Passive plasmonic demultiplexers using multimode interference. Journal of Lightwave Technology, 36(10), 1979-1984.

[14] Binfeng, Y., Guohua, H., Ruohu, Z., & Yiping, C. (2014). Design of a compact and high sensitive refractive index sensor base on metal-insulator-metal plasmonic Bragg grating. Optics Express, 22(23), 28662-28670.

- [15] Chen, L., Liu, Y., Yu, Z., Wu, D., Ma, R., Zhang, Y., & Ye, H. (2016). Numerical analysis of a near-infrared plasmonic refractive index sensor with high figure of merit based on a fillet cavity. Optics express, 24(9), 9975-9983.
- [16] Wang, G., Lu, H., & Liu, X. (2012). Trapping of surface plasmon waves in graded grating waveguide system. Applied Physics Letters, 101(1), 013111.
- [17] Li, D., Du, K., Liang, S., Zhang, W., & Mei, T. (2016). Wide band dispersion less slow light in hetero-MIM plasmonic waveguide. Optics express, 24(20), 22432-22437.
- [18] Yang, X., Hu, X., Chai, Z., Lu, C., Yang, H., & Gong, Q. (2014). Tunable ultracompact chip-integrated multichannel filter based on plasmon-induced transparencies. Applied Physics Letters, 104(22), 221114.
- [19] Yang, X., Hu, X., Yang, H., & Gong, Q. (2017). Ultracompact all-optical logic gates based on nonlinear plasmonic nanocavities. Nanophotonics, 6(1), 365.
- [20] Chai, Z., Hu, X., Yang, H., & Gong, Q. (2017). Chip-integrated all-optical diode based on nonlinear plasmonic nanocavities covered with multicomponent nanocomposite. Nanophotonics, 6(1), 329.
- [21] Hill, M. T., Marell, M., Leong, E. S., Smalbrugge, B., Zhu, Y., Sun, M., ... & Nötzel, R. (2009). Lasing in metal-insulator-metal sub-wavelength plasmonic waveguides. Optics express, 17(13), 11107-11112.
- [22] Yu, Z., Veronis, G., Fan, S., & Brongersma, M. L. (2008). Gain-induced switching in metal-dielectric-metal plasmonic waveguides. Applied Physics Letters, 92(4), 041117.
- [23] Zhang, Z., Yang, J., He, X., Han, Y., Zhang, J., Huang, J., ... & Xu, S. (2018). Active Enhancement of Slow Light Based on Plasmon-Induced Transparency with Gain Materials. Materials, 11(6), 941.
- [24] Lu, H., Gan, X., Mao, D., & Zhao, J. (2017). Graphene-supported manipulation of surface plasmon polaritons in metallic nanowaveguides. Photonics Research, 5(3), 162-167.
- [25] Chen, Z., Zhou, R., Wu, L., Yang, S., & Liu, D. (2018). Surface plasmon characteristics based on graphene-cavity-coupled waveguide system. Solid State Communications, 280, 50-55.
- [26] Song, C., Wang, J., Liu, D., Hu, Z. D., & Zhang, F. (2018). Wavelength-sensitive PIT-like double-layer graphene-based metal–dielectric–metal waveguide. Applied optics, 57(33), 9770-9776.
- [27] Li, Y., & Argyropoulos, C. (2018). Tunable nonlinear coherent perfect absorption with epsilon-near-zero plasmonic waveguides. Optics letters, 43(8), 1806-1809.
- [28] Huang, Y., Min, C., & Veronis, G. (2016). Broadband near total light absorption in non-PT-symmetric waveguide-cavity systems. Optics express, 24(19), 22219-22231.
- [29] Wuttig, M., Bhaskaran, H., & Taubner, T. (2017). Phase-change materials for non-volatile photonic applications. Nature Photonics, 11(8), 465.
- [30] Youngblood, N., Ríos, C., Gemo, E., Feldmann, J., Cheng, Z., Baldycheva, A., ... & Bhaskaran, H. (2019). Tunable Volatility of $Ge_2Sb_2Te_5$ in Integrated Photonics. Advanced Functional Materials, 29(11), 1807571.
- [31] Lankhorst, M. H., Ketelaars, B. W., & Wolters, R. A. (2005). Low-cost and nanoscale non-volatile memory concept for future silicon chips. Nature materials, 4(4), 347.
- [32] Ríos, C., Stegmaier, M., Hosseini, P., Wang, D., Scherer, T., Wright, C. D., ... & Pernice, W. H. (2015). Integrated all-photonic non-volatile multi-level memory. Nature Photonics, 9(11), 725.
- [33] Feldmann, J., Youngblood, N., Wright, C. D., Bhaskaran, H., & Pernice, W. H. P. (2019). All-optical spiking neurosynaptic networks with self-learning capabilities. Nature, 569(7755), 208.
- [34] Rudé, M., Pello, J., Simpson, R. E., Osmond, J., Roelkens, G., van der Tol, J. J., & Pruneri, V. (2013). Optical switching at 1.55 μm in silicon racetrack resonators using phase change materials. Applied Physics Letters, 103(14), 141119.
- [35] Zheng, J., Khanolkar, A., Xu, P., Colburn, S., Deshmukh, S., Myers, J., ... & Boechler, N. (2018). GST-on-silicon hybrid nanophotonic integrated circuits: a non-volatile quasi-continuously reprogrammable platform.

Optical Materials Express, 8(6), 1551-1561.

[36] Rios, C., Stegmaier, M., Cheng, Z., Youngblood, N., Wright, C. D., Pernice, W. H., & Bhaskaran, H. (2018). Controlled switching of phase-change materials by evanescent-field coupling in integrated photonics. Optical Materials Express, 8(9), 2455-2470.

[37] Wang, Q., Rogers, E. T., Gholipour, B., Wang, C. M., Yuan, G., Teng, J., & Zheludev, N. I. (2016). Optically reconfigurable metasurfaces and photonic devices based on phase change materials. Nature Photonics, 10(1), 60.

[38] Gholipour, B., Zhang, J., MacDonald, K. F., Hewak, D. W., & Zheludev, N. I. (2013). An all-optical, non-volatile, bidirectional, phase-change meta-switch. Advanced materials, 25(22), 3050-3054.

[39] Chen, Y. G., Kao, T. S., Ng, B., Li, X., Luo, X. G., Luk'Yanchuk, B., ... & Hong, M. H. (2013). Hybrid phase-change plasmonic crystals for active tuning of lattice resonances. Optics express, 21(11), 13691-13698.

[40] Palik, E. D. (Ed.). (1998). Handbook of optical constants of solids (Vol. 3). Academic press.

[41] Chu, C. H., Tseng, M. L., Chen, J., Wu, P. C., Chen, Y. H., Wang, H. C., ... & Tsai, D. P. (2016). Active dielectric metasurface based on phase-change medium. Laser & Photonics Reviews, 10(6), 986-994.

[42] Tian, J., Luo, H., Yang, Y., Ding, F., Qu, Y., Zhao, D., ... & Bozhevolnyi, S. I. (2019). Active control of anapole states by structuring the phase-change alloy Ge 2 Sb 2 Te 5. Nature communications, 10(1), 396.

[43] Bavil, M. A., Gao, L., & Sun, X. (2013). A compact nanoplasmonics filter and intersection structure based on utilizing a slot cavity and a Fabry–Perot resonator. Plasmonics, 8(2), 631-636.

[44] Cai, W., White, J. S., & Brongersma, M. L. (2009). Compact, high-speed and power-efficient electrooptic plasmonic modulators. Nano letters, 9(12), 4403-4411.

[45] Min, C., & Veronis, G. (2009). Absorption switches in metal-dielectric-metal plasmonic waveguides. Optics Express, 17(13), 10757-10766.

[46] Zhong, Z. J., Xu, Y., Lan, S., Dai, Q. F., & Wu, L. J. (2010). Sharp and asymmetric transmission response in metal-dielectric-metal plasmonic waveguides containing Kerr nonlinear media. Optics express, 18(1), 79-86.

[47] Piao, X., Yu, S., & Park, N. (2012). Control of Fano asymmetry in plasmon induced transparency and its application to plasmonic waveguide modulator. Optics express, 20(17), 18994-18999.

[48] Haddadpour, A., Nezhad, V. F., Yu, Z., & Veronis, G. (2016). Highly compact magneto-optical switches for metal-dielectric-metal plasmonic waveguides. Optics letters, 41(18), 4340-4343.

[49] Lin, X. S., & Huang, X. G. (2008). Tooth-shaped plasmonic waveguide filters with nanometeric sizes. Optics letters, 33(23), 2874-2876.

[50] Wang, T., Zhang, Y., Hong, Z., & Han, Z. (2014). Analogue of electromagnetically induced transparency in integrated plasmonics with radiative and subradiant resonators. Optics express, 22(18), 21529-21534.